\documentclass{raa}            

\usepackage{graphicx,times}             
\usepackage{natbib}
\usepackage{amssymb,amsmath}
\bibpunct{(}{)}{;}{a}{}{,}

\usepackage[a4paper=true,pagebackref=true]{hyperref}
\hypersetup{colorlinks = true, linkcolor = green, anchorcolor = red, citecolor = blue, filecolor = red, pagecolor = red, urlcolor = red}

\begin{document}

   \title{Research of the HIP~18856 binary system.}

   \volnopage{Vol.0 (20xx) No.0, 000--000}      
   \setcounter{page}{1}          

   \author{P. Efremova
      \inst{1}
   \and A. Mitrofanova
      \inst{2}
   \and V. Dyachenko
      \inst{2}
   \and A. Beskakotov
      \inst{2}
   \and A. Maksimov
      \inst{2}
   \and D. Rastegaev
      \inst{2}
   \and Yu. Balega
      \inst{2}
   \and S. Komarinsky
      \inst{2}
   }

   \institute{Kazan Federal University, Kazan 420008, Russian Federation \\
        \and
             Special Astrophysical Observatory of the Russian Academy of Sciences, Nizhnij Arkhyz, Russia 369167; {\it arishka670a@mail.ru} \\    
\vs\no
   {\small Received~~20xx month day; accepted~~2020~~March 21}}

\abstract{The results of the study of the binary HIP~18856 and construction of its orbit are presented. New observational data obtained at the BTA of SAO RAS in 2007-2019. Earlier \citet{cve16} constructed orbit for this system. However, it is based on 6 measurements, which cover a small part of the orbit. The positional parameters of the \textit{Hipparcos} \citep{hip} astrometric satellite, published speckle interferometric data \citep{mas99, bali06, bali07, hor12, hor17} and new ones were used in this study. Based on the new orbital parameters, the mass sum was calculated, and the physical parameters of the components were found. The obtained orbital and fundamental parameters were compared with the data from the study by \citet{cve16}. The comparison shows that the new orbital solution is better than the old one, since it fits new observational data accurately. Also based on the qualitative grade performed by \citet{wor83} new orbit was classified as "reliable", which means data covering more than half of the orbit with sufficient quantities of residuals of measurements.
}

   \authorrunning{P. Efremova, A. Mitrofanova, V. Dyachenko et al.}            
   \titlerunning{Research of the HIP~18856 binary system.}  

   \maketitle

%
%
\twocolumn
\section{Introduction}           
\label{sect:intro}

As known, the resolution of telescopes is limited by the influence of the atmosphere and the diffraction limit of the telescope. When observing objects with a long exposure, the images are blurred by an unsteady turbulent atmosphere that distorts the wavefront. An increase in resolution to the diffraction limit is possible due to the speckle interferometry method proposed by \citet{lab70}. With a short exposures of about 0.01-0.02 second, changes in interference patterns are low. Speckle interferometric images make it possible to observe the fine structure of objects, despite atmospheric distortions. 

One way to apply this method is to obtain positional parameters of binaries with low-mass components and thereafter fundamental parameters directly from observational data. The monitoring of such objects is carried out at a 6-m telescope by the group of high-resolution methods in astronomy of the Special Astrophysical Observatory of the Russian Academy of Sciences (SAO RAS). A study of one of such systems - HIP~18856 - is presented in this paper. Section \ref{sect:Lit} presents literature review, section \ref{sect:Obs} is devoted to the description of speckle interferometric observations and their reduction, section \ref{sect:Orb} - to the orbit construction and determination of fundamental parameters. The results are discussed in Section \ref{Dis}.

\section{Literature Review}
\label{sect:Lit}

The object in this study is the star HIP~18856 ($\alpha = 04^{h} 02^{m} 32.83^{s}$, $\delta = +06\degr 37' 52.08\arcsec$, $V_{mag} = 10.8^{m}$). It was resolved as a binary by the \textit{Hipparcos} astrometric satellite in 1991 \citep{hip}. Its binarity has been confirmed by speckle interferometric data obtained by different authors \citep{mas99, bali06, bali07, hor12, hor17}. Published positional parameters are presented in Table \ref{tab1}. For the system under study there is an orbital solution by \citet{cve16}, constructed using six measurements published previously in the literature. Measurements cover less than half of the orbit, and the resulting orbital parameters have low accuracy (Table \ref{tab2}). The paper by \citet{cve16} also gives the spectral types and masses of the components of HIP~18856 and their mass sum (Table \ref{tab3}). To calculate the parameters of the object, the authors used the \textit{Hipparcos} parallax and calculated the dynamical parallax $\pi_{dyn} = 13.24 \pm 0.28$ mas. \citet{mcd12} calculated the effective temperature of the object $T_{eff}=4800~K$, luminosity $L=0.34~L_{\odot}$ and determined the infrared excess $E_{IR}=1.181$, that probably indicates the existence of additional component.

\section{Observations and Data Reduction}
\label{sect:Obs}

The observations of HIP~18856 were carried out at the Big Telescope Alt-azimuth (BTA) of the SAO RAS (classical reflector, D = 6m, F = 24m) in 2007-2019. Speckle images were obtained using a speckle interferometer \citep{maks09} installed in the primary focus of the telescope. We used EMCCD's, which provides high sensitivity together with high time resolution: PhotonMAX-512B (untill 2010), Andor iXon+ X-3974 (2010-2014) and Andor iXon Ultra 897 (since 2015). The observational data are series of 1940 (until 2010) and 2000 short-exposures (with an exposure time of 20 ms) images. Speckle interferograms on December 13, 2008 were obtained using a $\times$20 micro-lens (field of view 3.7\arcsec) and a $\times$16 micro-lens (field of view 4.4\arcsec) on the other observational nights. Interference filters 600/40 nm and 800/100 nm were used. HIP~18856 observations were carried out under good and excellent weather conditions, with seeing 0.8\arcsec-2.0\arcsec. 

Power spectra, autocorrelation functions and reconstructed images, that allows for the determination of the true position of the secondary relative to the primary, are results of the reduction of speckle interferometric observations. The analysis of power spectra and autocorrelation functions was described by \citet{bali02} and \citet{pluz05}, image reconstruction was carried out using the bispectral analysis method \citep{lohm83}. Positional parameters and the magnitude differences between components are listed in Table \ref{tab1}. The columns correspond to: epoch of observations in fractions of the Besselian year, telescope, interference filter $\lambda$/$\Delta\lambda$, $\theta$ is the position angle, $\rho$ is the angular separation, $\Delta m$ - magnitude difference and reference. The accuracy of the parameters is due to the limiting angular resolution of the telescope, local atmospheric parameters at the time of observations and aberrations of the entire optical path from the mirror of telescope to the detector in the speckle interferometer. The actual measurement accuracy of $\Delta m$ is about 0.1 mag, and the table lists formal errors corresponding to the model selection method.

\begin{table*}
	\begin{center}
		\caption[]{HIP~18856 Positional Parameters and Magnitude Difference Between Components.}\label{tab1}
		\begin{tabular}{|c|c|c|c|c|c|c|}
			\hline\noalign{\smallskip}
Date & Telescope & $\lambda$/$\Delta\lambda$, nm & $\theta, \degr$ & $\rho$, mas & $\Delta m$, mag & Reference \\
			\hline\noalign{\smallskip}
1991.25	&	\textit{Hipparcos}	&		&	100.0	&	270	&	&	\cite{hip}	\\
1997.7182	&	2.1-m Otto Struve &	560/45	&	108.0	&	198	&	& \citet{mas99} \\
2000.8758	&	BTA	&	800/110	&	$117.2 \pm 0.6$	&	$162 \pm 2$	& $0.3 \pm 0.09$	&	\citet{bali06} \\	
2001.7530	&	BTA	&	800/110	&	$121.1 \pm 0.6$	&	$144 \pm 2$	& $0.19 \pm 0.34$	&	\citet{bali06} \\
2004.8214	&	BTA	&	800/110	&	$150.3 \pm 0.7$	&	$77 \pm 2$	& $0.28 \pm 0.05$	&	\citet{bali07} \\
2007.7372	&	BTA	&	600/40	&	$234.9 \pm 0.1$	&	$63 \pm 1$	&	$0.76 \pm 0.01$	& this work \\
2007.9776	&	BTA	&	600/40	&	$240.7 \pm 0.2$	&	$65 \pm 1$	&	$0.75 \pm 0.02$	& this work \\
2008.6996	&	3.5-m WIYN	&	698/39	&	74.3	&	75	&	0.43	&	\citet{hor12} \\
2008.9522	&	BTA	&	600/40	&	$259.7 \pm 0.2$	&	$76 \pm 1$	&	$2.35 \pm 0.03$	& this work \\
2009.9083	&	BTA	&	800/100	&	$279.9 \pm 0.1$	&	$71 \pm 1$	&	$0.97 \pm 0.04$	& this work \\
2011.6894	&	3.5-m WIYN &	692/40	&	151.5	&	50.2	&	0.35	&	\citet{hor17} \\
2011.6894	&	3.5-m WIYN &	880/50	&	159.5	&	54.0	&	0.68	&	\citet{hor17} \\
2011.9407	&	3.5-m WIYN &	692/40	&	339.3	&	55.7	&	0.69	&	\citet{hor17} \\
2011.9407	&	3.5-m WIYN &	562/40	&	343.8	&	52.8	&	0.05	&	\citet{hor17} \\
2011.9560	&	BTA	&	800/100	&	$344.2 \pm 0.1$	&	$51 \pm 1$	&	$0.52 \pm 0.01$	& this work \\
2015.8303	&	BTA	&	800/100	&	$58.4 \pm 0.1$	&	$127 \pm 1$	&	$0.50 \pm 0.01$	& this work \\
2016.8869	&	BTA	&	800/100	&	$64.6 \pm 0.1$	&	$151 \pm 1$	&	$0.53 \pm 0.02$	& this work \\
2017.7714	&	BTA	&	800/100	&	$67.2 \pm 0.1$	&	$170 \pm 1$	&	$0.62 \pm 0.02$	& this work \\
2018.0721	&	BTA	&	800/100	&	$68.7 \pm 0.1$	&	$177 \pm 1$	&	$0.56 \pm 0.01$	& this work \\
2019.0494	&	BTA	&	800/100	&	$72.0 \pm 0.1$	&	$195 \pm 1$	&	$0.56 \pm 0.01$	& this work \\
			\noalign{\smallskip}\hline
		\end{tabular}
	\end{center}
\end{table*}

\section{Orbit Construction}
\label{sect:Orb}

The orbit of HIP~18856 was constructed based on the positional parameters from Table \ref{tab1}: a preliminary orbital solution was found using the Monet method \citep{mon77}, and the final orbit - using the ORBIT software package \citep{tok92}. During the construction of the orbit, we found that the positions of the published measurements 2008.6996 \citep{hor12} and 2011.6894 (both measurements) \citep{hor17} must be changed by $180\degr$.

\begin{figure}
 	\centering
 	\includegraphics[width=8cm, angle=0]{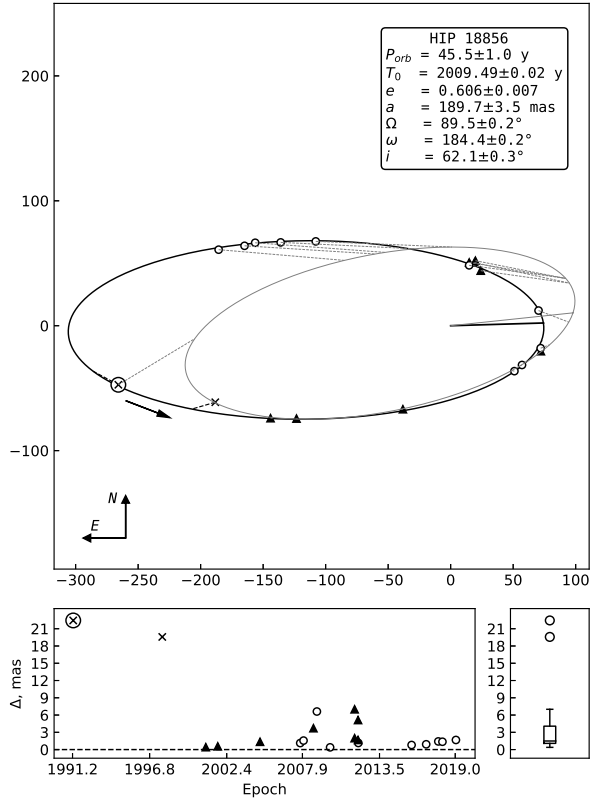}
	\caption{Orbital solutions for HIP~18856.}
	\label{fig1}
\end{figure}

Figure \ref{fig1} shows a comparison of two orbits: the orbit by \citet{cve16} is marked with gray, improved orbit marked with black; triangles correspond to published data; open circles - new data; crosses - data with large values of residuals; a point placed in a large circle is the first measurement for the system; the arrow shows the direction of motion of the secondary. $\Delta$ - residuals showing the difference between the observed and modelled value; the dashed line on the residuals plot indicates the orbital solution.

The average estimates of the residuals were $2.2\degr$ for $\theta$ and 6 mas for $\rho$. After excluding outliers (in Figure \ref{fig1}, these are marked with crosses), discrepancies for $\rho$ and $\theta$ were 1.3 mas and $2.1\degr$, which indicates a good fitting of observational data with a new orbital solution. The orbits shown in Figure \ref{fig1} are very different from each other, the published solution does not fit observational data obtained after 2008. It is worth noting that \citet{cve16} used the measurement 2008.6996 \citep{hor12}, the positional angle of which should be changed by $180\degr$. The orbital parameters of HIP~18856 obtained in this work and published by \citet{cve16} are presented in Table \ref{tab2}. The columns shows: $P_{orb}$ is the orbital period of the system, $T_{0}$ is the epoch of passing the periastron, $e$ is the eccentricity of the orbit, $a$ is the semimajor axis, $\Omega$ is the longitude of the ascending node, $\omega$ is the argument of the periastron, $i$ is the inclination of the orbit and the reference to the publication. Thus, the orbit of HIP~18856 was improved due to long-term speckle interferometric observations at the BTA of the SAO RAS (in particular, the orbital period and the semimajor axis were significantly increased).

\begin{table*}
	\begin{center}
		\caption[]{Orbital Parameters of HIP~18856.}\label{tab2}
		\begin{tabular}{|c|c|c|c|c|c|c|c|}
			\hline\noalign{\smallskip}
$P_{orb}$, year & $T_{0}$, year & $e$ & $a$, mas & $\Omega$, $\degr$ &  $\omega$, $\degr$ & $i$, $\degr$ & Reference \\
			\hline\noalign{\smallskip}
35.441	&	2010.303	&	0.366		&	159.0		&	102.0		&	166.1		&	64.9		&	\citet{cve16}	\\ 
$\pm 0.517$	&	$\pm 0.410$	&	$\pm 0.115$	&	$\pm 1.2$	&	$\pm 3.3$	&	$\pm 9.5$	&	$\pm 2.5$	&			\\ 
\hline\noalign{\smallskip}
45.5		&	2009.49	&	0.606		&	189.7		&	89.5		&	184.4		&	62.1		& this \\
$\pm 1.0$	&	$\pm 0.02$	&	$\pm 0.007$	&	$\pm 3.5$ 	&	$\pm 0.2$	&	$\pm 0.2$	&	$\pm 0.3$	& work\\
			\noalign{\smallskip}\hline
		\end{tabular}
	\end{center}
\end{table*}

The fundamental parameters of the components and the system were determined by two independent methods: using the Kepler’s third law (Equations \ref{eq1} and \ref{eq2}) and the Pogson's relation. Apparent magnitude of the object in V-band from the SIMBAD database, the average magnitude difference in the 600/40 filter (assuming this value is close to the value in the V-band) and the table of fundamental parameters of the main sequence stars published by \citet{pec13} were used in the second method. It should be noted that the characteristics were determined for the \textit{Hipparcos} \citep{hippi} and \textit{Gaia} \citep{gapi} parallaxes, which is an additional indicator of the correctness of the obtained orbit.

\begin{equation}
\sum \mathfrak{M}=\frac{(a/\pi)^{3}}{P_{orb}^{2}}
\label{eq1}
\end{equation}

\begin{equation}
\begin{gathered}
\sigma(\mathfrak{M})=\\
\sqrt{\frac{9(\sigma_{\pi})^2}{\pi^2}+\frac{9(\sigma_{a})^2}{a^2}+\frac{4(\sigma_{P_{orb}})^2}{P_{orb}^2}}*\mathfrak{M}
\end{gathered}
\label{eq2}
\end{equation}

The fundamental parameters determined by the two methods for the parallaxes of the \textit{Hipparcos} ($\pi_{\textit{Hip}} = 14.17 \pm 2.79$ mas) and \textit{Gaia} ($\pi_{\textit{Gaia}} = 13.4777 \pm 0.2577$ mas) missions and obtained by \citet{cve16} are presented in Table \ref{tab3}. It should be noted, that the errors of parameters are primarily due to errors of parallax. The resulting mass sums and the characteristics of the components are in good agreement with each other, which confirms the correctness of the constructed orbit. 

\begin{table*}
	\begin{center}
		\caption[]{Fundamental Parameters of HIP~18856 and System Components.}\label{tab3}
		\begin{tabular}{|c|c|c|c|c|c|c|c|c|}
			\hline\noalign{\smallskip}
& Parallax & $M_{V,A}$, m & $Sp_{A}$ & $\mathfrak{M}_{A}$, $\mathfrak{M}_{\odot}$ & $M_{V,B}$, m &  $Sp_{B}$ & $\mathfrak{M}_{B}$, $\mathfrak{M}_{\odot}$ & $\sum \mathfrak{M}$, $\mathfrak{M}_{\odot}$ \\
			\hline\noalign{\smallskip}
\citet{cve16} & \textit{Hipparcos}	&	$6.43 \pm 0.48$	&	K2	&	0.81	&	$6.87 \pm 0.72$	&	K3	&	0.79	&	$1.12 \pm 0.67$	\\ 
\hline\noalign{\smallskip}
this &	\textit{Hipparcos}	&	$6.99 \pm 0.06$	&	K4	&	0.72	&	$7.75 \pm 0.06$	&	K6	&	0.65	&	$1.16 \pm 0.69$	\\ \cline{2-9}
work 	& \textit{Gaia}	&	$6.89 \pm 0.06$	&	K4	&	0.72 	&	$7.64 \pm 0.06$	&	K5.5	&	0.66	& $1.35 \pm 0.12$\\
			\noalign{\smallskip}\hline
		\end{tabular}
	\end{center}
\end{table*}

\section{Discussion}
\label{Dis}

Long-term monitoring of HIP~18856 was carried out in the group of high-resolution methods in astronomy of the SAO RAS at the 6-m telescope from 2007 to 2019, which made it possible to improve the orbit of the system. The positional parameters obtained in this study doubled the number of available published observational data. The new orbit fits speckle interferometric observations obtained after 2008 better, than the previously one found by \citet{cve16}. The errors of mass sum are 59\% when calculating using the parallax of the \textit{Hipparcos} mission and 9\% using the parallax of the \textit{Gaia} mission. This low accuracy of the mass sum using the \textit{Hipparcos} parallax is due to its own low accuracy (it should be noted that the third power of this value is used in Equation \ref{eq1}). The fundamental parameters determined by two independent methods and using two parallaxes are in good agreement with each other. For a qualitative analysis of the obtained orbital solution, we used the classification of orbits by \citet{wor83}. This classification is based on number of observational data, their residuals and orbit coverage. Orbits are graded from 1 ("Definitive") to 5 ("Indeterminate"). The new orbit of HIP~18856 is "reliable" - this means that the measurements cover about half of the orbit with sufficient values of residuals of measurements. Monitoring of HIP 18856, carried out for approximately 28 years (starting with the Hipparcos mission), made it possible to construct an accurate orbit, that is not always available for objects with longer orbital periods.
	
This study shows the importance of long-term monitoring of speckle interferometric binaries. Despite the fact that orbits have already been constructed for many such systems, they are not always correct. Especially if these orbital solutions are obtained using a small amount of observational data. The use of new observational data minimizes errors in orbital parameters. Additionally high accuracy of the obtained parameters of speckle-interferometric binaries will reveal new relationships between the characteristics of systems and their components. Such studies will not only improve our knowledge of the fundamental parameters of binaries, but also reveal inaccuracies in data from space missions.

\begin{acknowledgements}
The reported study was funded by RFBR, project number 20-32-70120. The work was performed as part of the government contract of the SAO RAS approved by the Ministry of Science and Higher Education of the Russian Federation. This work has made use of data from the European Space Agency (ESA) mission \textit{Gaia} (\url{https://www.cosmos.esa.int/gaia}), processed by the \textit{Gaia} Data Processing and Analysis Consortium (DPAC, \url{https://www.cosmos.esa.int/web/gaia/dpac/consortium}). Funding for the DPAC has been provided by national institutions, in particular the institutions participating in the \textit{ Gaia} Multilateral Agreement. This research has made use of the SIMBAD database, operated at CDS, Strasbourg, France.
\end{acknowledgements}

\bibliographystyle{raa}
\bibliography{18856}

\label{lastpage}

\end{document}